%% file: main.tex
\documentclass[journal]{vgtc}

\onlineid{1171}

\vgtccategory{Research}

\title{Revisiting Channel Effectiveness: \\A Multi-Dimensional Evaluation with Primitive Visual Stimuli}

\author{%
    \authororcid{Soohyun Lee}{0000-0002-3075-3981},
    \authororcid{Seokhyeon Park}{0009-0003-1685-4027},
    \authororcid{Minsuk Chang}{0009-0007-5088-8991}, and
    \authororcid{Jinwook Seo}{0000-0002-7734-822X}
}

\authorfooter{
    \item
        Soohyun Lee, Seokhyeon Park, and Jinwook Seo are with Seoul National University.
        E-mail: \{shlee, shpark\}@hcil.snu.ac.kr; jseo@snu.ac.kr.%
    \item
        Minsuk Chang is with Georgia Institute of Technology.
        E-mail: minsuk@gatech.edu.%
}

\abstract{\input{sections/00_abstract}}

\keywords{Visual channels, graphical perception, channel effectiveness, primitive visual stimuli}

\teaser{
  \centering
  \includegraphics[width=0.8\linewidth, alt={Overview of the study. Left: the seven visual channels (position, length, tilt, area, curvature, luminance, saturation) rendered as primitive stimuli. Right: the four perceptual tasks (accuracy, discriminability, separability, and pop-out), each illustrated with an example stimulus and its response format.}]{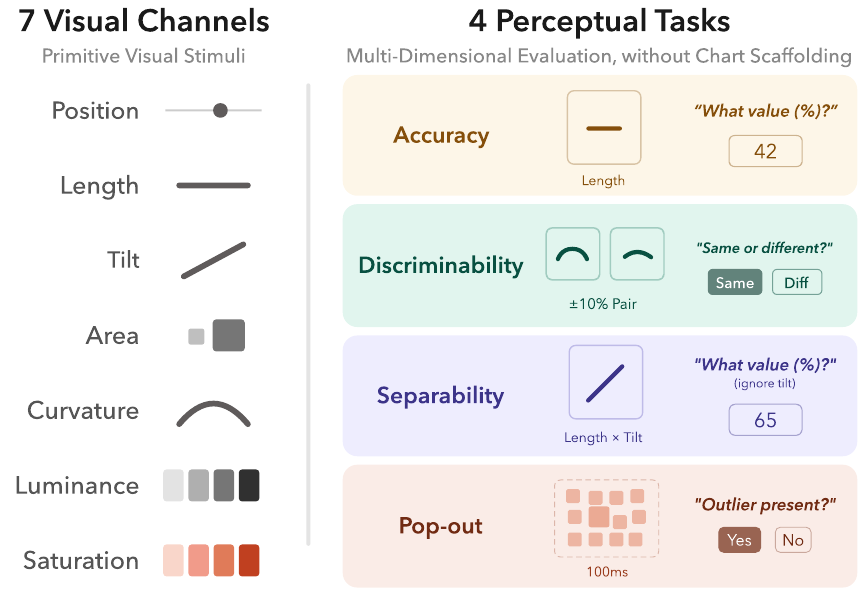}
  \caption{%
    We evaluate seven visual channels (left) across four perceptual tasks: accuracy, discriminability, separability, and pop-out.
    Every task uses primitive stimuli stripped of chart scaffolding.
    Each row shows an example stimulus, the question posed to participants, and a sample response (right).
  }
  \label{fig:teaser}
}

\graphicspath{{figs/}{figures/}{pictures/}{images/}{./}}

\usepackage{mathptmx}
\usepackage{amsmath,amssymb}
\usepackage{booktabs}
\usepackage{array}
\usepackage{multirow}
\usepackage{colortbl}
\usepackage{xcolor}
\usepackage{diagbox}
\usepackage{pifont}
\usepackage{stfloats}
\newcommand{\cmark}{\ding{51}}

\begin{document}


\maketitle

\section{Introduction}

\input{sections/01_intro}
\input{sections/02_relwork}
\input{sections/03_method}
\input{sections/04_Accuracy}
\input{sections/05_Discriminability}
\input{sections/06_Separability}
\input{sections/07_Popout}
\input{sections/08_design_implications}
\input{sections/09_result}

\acknowledgments{%
    This work was supported by the National Research Foundation of Korea (NRF) grant funded by the Korean government (MSIT) (No. NRF-2023R1A2C2005209), the Institute of Information \& Communications Technology Planning \& Evaluation (IITP) grant funded by the Korean government (MSIT) [No. RS-2021-II211343, Artificial Intelligence Graduate School Program (Seoul National University)], the AI Seoul Tech Research Support Program of the Seoul Future Foundation, and the SNU-Global Excellence Research Center establishment project.
    The ICT at Seoul National University provided research facilities for this study.
}

\bibliographystyle{abbrv-doi-hyperref}
\bibliography{ref}
\clearpage
\appendix

\input{sections/10_supplementary}

\end{document}

%% file: sections/00_abstract.tex
Established channel effectiveness rankings primarily assess magnitude estimation accuracy in complete chart contexts, often neglecting other perceptual tasks such as discriminability, separability, and pop-out.
To address this gap, we conducted crowdsourced experiments on seven core visual channels (position, length, tilt, area, curvature, luminance, and saturation) using primitive visual stimuli, a set of visual marks without chart-specific scaffolding to isolate channel-level variation.
We evaluated these channels across four perceptual tasks (accuracy, discriminability, separability, and pop-out) and found that channel effectiveness is fundamentally multi-dimensional, with rankings shifting substantially across tasks.
For instance, while spatial channels maintain an overall advantage, accuracy depends strongly on whether a fixed spatial anchor is available.
Discriminability varies dramatically across channels and value ranges, a pattern we formalized with a novel Anchored Harmonic Weber model.
Pairwise channel interactions are often strongly asymmetric.
Finally, we identify a dissociation between estimation accuracy and preattentive detection: length shows only moderate detection effectiveness despite top-tier accuracy, while area achieves the highest detection rates despite poor quantitative accuracy, though the latter advantage may partly reflect stimulus-level cues.
We synthesize these findings into a scenario-driven perspective for context-sensitive channel selection.

%% file: sections/01_intro.tex
Visual channels such as position, length, tilt, area, and color are the fundamental building blocks of data visualization.
Cleveland and McGill's foundational work~\cite{GraphicalPerception}, later corroborated by Heer and Bostock~\cite{channelMturk}, established influential rankings showing that spatial channels (position, length) enable more accurate reading than others (angle, area, color).
This hierarchy has shaped design guidelines and automated visualization systems~\cite{automating} for decades.

However, this knowledge faces critical limitations in modern visualization design.
First, classic experiments tested channels within complete chart frameworks (bar charts with axes, scatter plots with gridlines), making it difficult to separate channel behavior from the perceptual scaffolding  these chart elements provide~\cite{GraphicalPerception}.
For instance, the perceptual accuracy of a pie chart varies depending on whether viewers rely on angle versus area~\cite{skau2016arcs, kiesel2025rose}, and radial layouts produce conflicting outcomes depending on the channel used~\cite{waldner2019comparison}.
Furthermore, existing work focuses primarily on accuracy while neglecting other essential perceptual tasks: discriminability (detecting small differences), separability (combining multiple channels), and pop-out (rapidly identifying outliers)~\cite{franconeri2021science, munzner}.

We address these gaps by testing seven channels (position, length, tilt, area, curvature, luminance, saturation) across four tasks (accuracy, discriminability, separability, and pop-out).
To do so, we evaluate channels using primitive visual stimuli, visual marks stripped of chart-specific scaffolding (e.g., a single line segment within a shared display frame), to isolate channel-level variation.

Our results both confirm and challenge established knowledge on channel effectiveness.
Spatial channels retain their accuracy advantage, consistent with classic rankings, but cross-task profiles reveal striking dissociations, most notably, area performs poorly in accuracy yet achieves the highest detection rate in our pop-out setting.
Such reversals show that channels sidelined by accuracy-centric rankings play a central role once performance is evaluated across multiple tasks.

Building on this multi-dimensional view, we use an equivalence margin and effect sizes to distinguish ranking differences that are practically meaningful from those that are negligible (for instance, tilt's accuracy is statistically equivalent to single position).
We synthesize these results into a scenario-driven perspective that maps each task to a recurring design dimension, enabling context-sensitive encoding choices beyond traditional chart conventions.

Our main contributions are:
\begin{enumerate} 
    \item A systematic evaluation of seven visual channels across four perceptual tasks (accuracy, discriminability, separability, and pop-out), with channel subsets and pairings adapted per task, using primitive visual stimuli that isolate channel-level variation from chart-specific scaffolding.
    \item Structured evidence that channel effectiveness is multi-dimensional: channel rankings shift substantially across the four tasks, and pairwise channel interactions exhibit pronounced asymmetries, demonstrating that no single ordering captures perceptual performance comprehensively.
    \item A scenario-driven perspective that organizes these findings into four recurring design dimensions (task stakes, data granularity, background interference, and response time) to guide context-sensitive encoding choices.
    \item Evidence of a dissociation between estimation accuracy and preattentive detection: length shows only moderate pop-out despite top-tier accuracy, while area achieves the highest detection rate despite poor quantitative accuracy (though area's advantage may partly reflect stimulus-level cues).
\end{enumerate}

%% file: sections/02_relwork.tex
\section{Related Work}

\subsection{Graphical Perception in Visualization Contexts 
}

Cleveland and McGill~\cite{GraphicalPerception} compared bar, pie, and scatter plots and showed that position and length enable lower error than angle or area in complete charts, and Mackinlay~\cite{automating} codified this ranking into automated design heuristics, tying perceptual accuracy to chart form rather than to the underlying channel. 
These channel taxonomies trace back to Bertin's visual variables~\cite{monmonier1985semiology}.
Heer and Bostock~\cite{channelMturk} confirmed these findings via crowdsourcing, yet their stimuli retained extensive scaffolding (e.g., axes, tick marks, gridlines, and legends).
Such scaffolding can inflate or depress accuracy. Axes simplify height estimation, while adjacent bubbles may distract or artificially simplify the task.
More recent computational work has applied CNNs to chart-level graphical perception~\cite{cnnPerception}, probed the sensitivity of vision models to isolated channels~\cite{soohyun}, and disentangled visual reading from factual priors in LVLMs' visualization literacy~\cite{lee2026disentangling}, but none of these studies combined human observers with context-free, single-channel stimuli.

More broadly, recent work has questioned whether a single channel ranking can serve all visualization tasks, showing that channel effectiveness depends on perceptual task and data distribution~\cite{McColeman2022, KimHeer2018, Saket2019}.
These findings motivate our multi-task evaluation rather than relying on a single performance dimension.

\subsection{Perceptual Foundations of Psychophysics and Channel Interactions}

Psychophysical theory links physical and perceived magnitude.
\textit{Stevens' power law} predicts near-linear growth for length but compressive curves for brightness and area~\cite{powerlaw}, whereas the \textit{Weber--Fechner} relation formalizes JND thresholds that underpin color-scale design~\cite{weberslaw}.

Subsequent work highlights categorical anchors.
Orientation adaptation studies show that $0^\circ$, $45^\circ$, and $90^\circ$ act as attractors, biasing estimates toward canonical angles~\cite{tiltaccuracy, tiltaccuracy2, tiltaccuracy_vertical}.
In color perception, Szafir derived difference functions for luminance and hue that inform palette design~\cite{SzafirColorDiscriminability}, and Bartram et al. demonstrated that grid-line color together with transparency jointly affect detection thresholds for overlaid grids~\cite{ColorTransparancy}.
These efforts reveal that perception is highly non-linear, even for single channels.
Yet systematic validation across the palette of encodings, particularly saturation and tilt, remains limited.

Beyond single-channel perception, preattentive processing (rapid, parallel detection of salient features before focused attention) is central to visualization design.
Feature Integration Theory~\cite{Treisman1980} established that basic features are registered in parallel while conjunctions require serial search~\cite{wolfe2004attributes}, and Healey and Enns identified which visual properties support rapid detection in visualization contexts~\cite{preattentive}.
Yet most pop-out studies in the visualization literature test features embedded in multi-element displays, leaving open whether the same preattentive hierarchies hold when chart-level scaffolding is absent.

When multiple channels are combined, the question of independence arises.
Garner's separable--integral framework~\cite{Garner1974} distinguishes channels that can be attended independently from those that interfere during perceptual judgment.
Smart and Szafir~\cite{SmartSzafir2019} measured the separability of shape, size, and color in scatterplots, finding significant interactions that violate simple independence assumptions.
However, systematic pairwise separability testing across a broader channel set remains lacking, particularly for geometric channels such as tilt and curvature, which have received limited empirical attention in visualization.
Our study addresses these gaps by unifying power-law fitting, Weber-fraction analysis, pairwise separability testing, and preattentive detection within a single crowdsourced protocol.

\subsection{Primitive-Stimulus and Single-Channel Experiments}

Recently, researchers have begun stripping chart context from the stimuli to conduct more precisely controlled experiments.
Szafir fit JND curves from single color swatches against a background~\cite{SzafirColorDiscriminability}, Veras and Collins manipulated exactly one glyph attribute per trial to compare human similarity ratings against an image-based discriminability metric~\cite{VerasDiscriminability}, and Demiralp et al. learned perceptual kernels from crowdsourced judgments over visualization stimuli~\cite{demiralp2014learning}.
Bartram et al. removed data marks entirely, varying only grid-line color and transparency to establish detection thresholds~\cite{ColorTransparancy}, while Bearfield et al. used unlabeled two- or three-bar arrays and uncovered length-comparison biases even under this reduced context~\cite{BearfieldGrouping}.

While each of these studies reduces stimuli, most retain contextual or competing elements, such as multiple glyphs, background grids, or adjacent bars, that can redirect attention.
They also tend to focus on a single channel and report either magnitude error or threshold sensitivity, but rarely both, leaving channels such as saturation and tilt untested under fully isolated conditions.

We extend this paradigm. A single mark varies along one channel with all other attributes fixed, letting us measure magnitude-estimation accuracy and fine-grained discriminability from the same observers.

%% file: sections/03_method.tex
\section{Method}
    
We conducted crowdsourced experiments with primitive visual stimuli to characterize channel behavior in a controlled setting that removes chart-specific scaffolding while retaining a shared display frame, testing accuracy (estimation precision), discriminability (just-noticeable differences), separability (cross-channel interference), and pop-out across seven visual channels.
We refer to these four conditions as tasks because each corresponds to a distinct perceptual operation that arises in real-world chart reading.
Figure~\ref{fig:teaser} provides an overview of the experimental framework and task structure across these four tasks.

\subsection{Stimuli Design}
\label{sec:stimuli_design}

Each stimulus consisted of exactly one visual mark (e.g., a line segment or filled square) that varied along a single encoding channel, while all other attributes remained fixed at default values.
All stimuli were displayed on a white background within a square canvas of 500$\times$500\,px bounded by a 1\,px black border. The canvas provides a shared display frame and bounds the drawable extent of the length and area channels; position is referenced to its line segment, while the color and angular ranges are established by the pre-block reference below.
We use \emph{primitive} to denote deconstructed marks that remove chart-specific scaffolding (axes, gridlines, labels, legends, adjacent marks) to isolate channel-level variation, retaining only the canvas boundary as a shared frame.
Isolating individual visual channels for controlled perceptual measurement has precedent in visualization research~\cite{bezerianos2012perception, soohyun}.
Before each experiment block, participants viewed a reference display showing the full range of the target channel (e.g., a color bar spanning 0\% to 100\% for luminance and saturation), ensuring comparable task references across spatial and chromatic channels.
Because every channel is judged relative to the on-screen canvas or the pre-block reference rather than in absolute physical units, we did not perform per-display calibration, as is standard for crowdsourced studies~\cite{channelMturk}. Variation in display hardware and viewing conditions therefore remains a limitation, particularly for the color channels.

We examined seven channels: position, length, tilt, area, curvature, luminance, and saturation, spanning the major magnitude channels emphasized in visualization effectiveness frameworks such as Munzner's~\cite{munzner}.
Position was further subdivided into three reference-frame variants detailed in Section~\ref{sec:accuracy}.
Hue and shape were included only in the pop-out experiment because the other three tasks require ordinal magnitude estimation on a bounded linear scale, which they lack without imposing an arbitrary origin (see Section~\ref{sec:popout}).

Metric channels (length, position, area, luminance, saturation) used integer steps from 0 to 100, while tilt and curvature used $0^\circ$--$180^\circ$ in $1^\circ$ increments.
For area, stimuli were filled squares spanning the smallest to largest shape allowed within the display region.
For curvature, the two endpoints matched the length stimulus, while the connecting path formed a circular arc whose central angle ranged from 0$^\circ$ (straight) to 180$^\circ$ (semicircle).

Default stimulus parameters (single centered line; 50\% length, $0^\circ$ tilt, 50\% luminance, 0\% saturation, $0^\circ$ curvature, red hue ($0^\circ$)) provided a neutral baseline so that non-target channels would not distract from the channel under investigation.
The color stimuli fix the HSL hue at the default red ($0^\circ$); luminance follows the HSL lightness axis ($0\%=$ black to $100\%=$ white) and saturation the HSL saturation axis ($0\%=$ gray to $100\%=$ fully saturated red), each mapping its $0$--$100\%$ value directly onto that coordinate.

\subsection{Participants and Procedure}

All experiments were implemented and deployed using the reVISit framework~\cite{ding2023revisit, cutler2026revisit}\footnote{\url{https://revisit.dev/}}, an open-source platform for conducting scalable online visualization studies.
Each stimulus was implemented as a React component within the reVISit environment, enabling programmatic generation of primitive visual stimuli with precise control over channel parameters.
We leveraged reVISit's built-in support for trial randomization, response capture, and integration with Amazon Mechanical Turk (MTurk) for participant recruitment and routing.
Our study was within-subjects, and we recruited 119 participants through MTurk, limiting eligibility to workers with at least a 97\% approval rate and 5000 or more approved HITs.
Across the core participant pool, 58\% identified as male and 42\% as female, the mean age was 36.9 years ($\pm 12.1$ SD), and 99\% reported at least some college education.
Attention-check trials presented stimuli with unambiguous ground-truth values (e.g., clearly extreme values such as 0 or 100 for metric channels, or maximally distinct stimulus pairs for discriminability trials); 14 participants who responded incorrectly on these trials were excluded from all analyses.
All 105 core participants completed all four tasks (Accuracy, Separability, Discriminability, and Pop-out), and we recruited an additional 45 participants for Discriminability to obtain more fine-grained samples.

The study took approximately 12 minutes in total, where participants provided informed consent, completed the assigned task blocks, and reported demographics (age range, gender, and education level).
Before the main task blocks, participants completed one or two practice rounds per task to ensure they understood the response formats.
Both task-block order and channel presentation order were fully randomized per participant to mitigate learning and fatigue effects.
For each main trial, participants typed numeric values (Accuracy and Separability), chose ``Same/Different'' (Discriminability), or indicated outlier presence (Pop-out).
Trials were self-paced with no explicit time limit; the pop-out task additionally enforced a 100\,ms stimulus exposure in a guided, uniform-distractor design, as described in Section~\ref{sec:popout}.
Trial counts varied by task: Accuracy (2), Discriminability (10 for the core participants; 32 for the additional participants), Pop-out (3), and all tested primary-secondary combinations in Separability; Accuracy and Pop-out counts were kept low to prioritize breadth across $7$ channels $\times$ $4$ tasks, whereas Discriminability required denser sampling to estimate JND contours.
Participants were compensated \$2 upon completion of all assigned trials, approximately \$10 per hour.
The Seoul National University IRB approved the protocol (No.~2503/004-016).
All four task blocks can be run at {\footnotesize\url{https://graphical-perception.github.io/Primitive-Study/}}.

%% file: sections/04_Accuracy.tex
\section{Accuracy}
\label{sec:accuracy}

\subsection{Task Design}
In each trial, participants performed a magnitude-estimation task, observing a single primitive stimulus and estimating the target channel's magnitude on its natural scale (0--100\% for metric channels; $0^\circ$--$180^\circ$ for tilt and curvature).

We tested seven channels: position, length, tilt, area, luminance, saturation, and curvature.
For the \textbf{position} channel, we designed three variants to probe how reference frames affect magnitude judgment:
\begin{itemize}
    \item \textit{Single position:} A single line segment is presented with a marked point; participants estimate what percentage from the start endpoint the mark appears.
    Both horizontal and vertical reference lines were used (start endpoint at the left or bottom, respectively).
    Note that this task is functionally equivalent to judging length from a fixed anchor.
    \item \textit{Position comparison (aligned):} Two identical, parallel line segments are displayed, each with a marked point, sharing a common baseline (their start endpoints lie on the same line).
    Participants report the absolute difference (in percentage points) between the relative positions of the two marked points along their segments.
    \item \textit{Position comparison (unaligned):} The same two parallel segments are presented, but each is translated in the opposite direction along its own length axis, so their start endpoints no longer share a common baseline.
    Participants again estimate the absolute difference (in percentage points) between the relative positions of the marked points.
\end{itemize}
Stimulus images for all variants and channels appear in Figure~\ref{fig:stim}.
For the \textbf{length} channel, participants judged a center-aligned horizontal line segment that extended bidirectionally from its midpoint, estimating its length as a percentage of the maximum possible length (0--100\%).

We obtained data from 105 participants, with two responses per participant for each channel condition (210 responses per channel).
For each trial, a value (or pair of values) was randomly selected within the channel's domain, with all true percentages uniformly distributed across the possible range.

\subsection{Analysis}

To analyze perceptual accuracy, we compared each participant's response ($P$) to the ground-truth value ($S$), both normalized to $[0,1]$ by the channel maximum $S_\mathrm{max}$ (e.g., 100 for length, 180 for tilt).
We modeled the response as a power-law transform $S^\alpha$, with $\alpha$ either fixed at $1$ (baseline linear model) or fit per channel by a fine grid search over $\alpha\in[0.25,4]$ minimizing the mean log-error. The fitted exponents appear in Figure~\ref{fig:accuracy_error}.
\vspace{-1em}
\begin{equation}
    \text{Mean log-error} = \frac{1}{N}\sum_{i=1}^{N} \log_2\!\bigl(|P_i - S_i^\alpha| + \tfrac{1}{8}\bigr)
    \label{eq:mean_log_error}
\vspace{-1em}
\end{equation}
The additive $\tfrac{1}{8}$ offset, adapted from Cleveland and McGill~\cite{GraphicalPerception}, prevents the logarithm from diverging at zero error; its effect is examined in Supplementary~\ref{sup:offset}.
All pairwise comparisons used paired $t$-tests on per-participant means, with Benjamini--Hochberg FDR correction (FDR level 0.05, 36 comparisons across nine conditions and seven channels, with position subdivided into three reference-frame variants).
We chose Benjamini--Hochberg over Bonferroni because the large number of comparisons makes family-wise error control overly conservative, whereas BH maintains statistical power while controlling the expected proportion of false discoveries~\cite{benjamini1995controlling}.
We checked the normality of the paired differences with Shapiro--Wilk tests and cross-checked headline comparisons with Wilcoxon signed-rank tests (Supplementary~\ref{sup:normality}).
For channels hypothesized to perform equivalently, we additionally applied two one-sided tests (TOST) to assess whether observed differences fell within a pre-specified equivalence margin.

\subsection{Results}

\begin{figure}
    \centering
    \includegraphics[width=\linewidth, alt={Dot plot of accuracy by channel. Single position and tilt show the lowest error, followed by length, aligned position, curvature, saturation, unaligned position, area, and luminance. Power-law correction markedly improves area (fitted exponent 0.424) but barely changes the color channels.}]{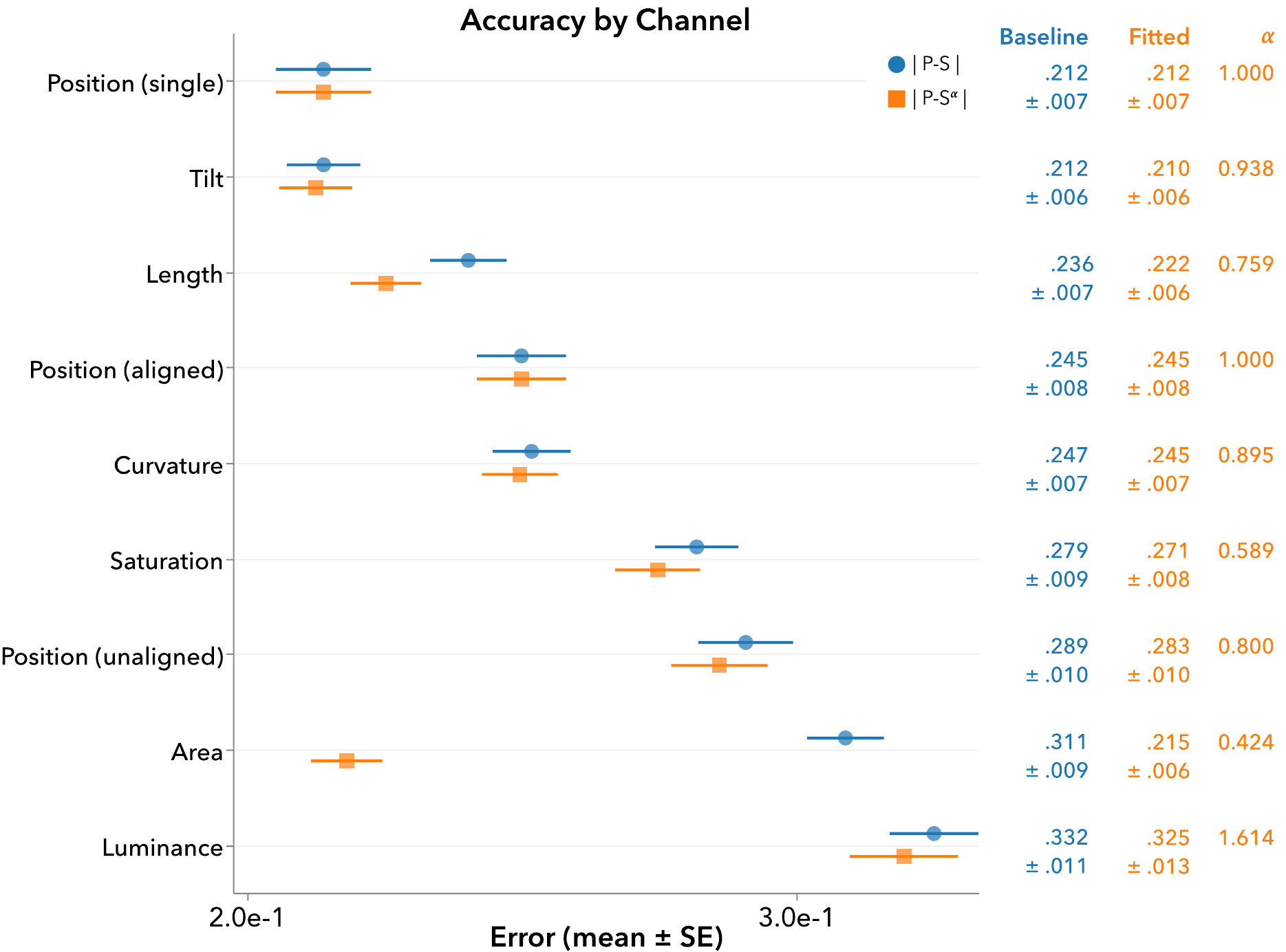}
    \caption{
        Per-channel accuracy: baseline (\textcolor[HTML]{1F77B4}{blue}, $\alpha{=}1$) and power-law corrected (\textcolor[HTML]{E67E22}{orange}, optimal $\alpha$).
        Markers show $2^{m}$, the back-transformed mean log-error, with horizontal bars denoting $\pm 1$ SE; lower is more accurate. Values annotated at right.
    }
    \label{fig:accuracy_error}
\vspace{-1em}
\end{figure}

Figure~\ref{fig:accuracy_error} summarizes the mean log-error for each channel, comparing the baseline linear model (blue) and best-fit power-law model (orange).
Lower values indicate higher perceptual accuracy.
The reported comparisons below survived FDR correction.
Cross-channel comparisons are shaped in part by the display frame, which offers stronger within-trial reference cues to spatial channels than to color (Section~\ref{sec:synthesis}).

\vspace{1mm}
\textbf{Position (single)} achieved the highest accuracy, significantly outperforming \textbf{length} (paired $t$-test, $p = 0.009$, Hedges' $g = 0.292$, FDR-corrected).
Unlike single position, which is read from a fixed end anchor, the length stimulus extends bidirectionally from its midpoint.
This bidirectional extension produced systematic overestimation ($\alpha = 0.759$), particularly at smaller values, consistent with the cognitive cost of integrating two segments from a center point.

\textbf{Position (aligned)} condition, where participants compared positions on two horizontally aligned lines, performed worse than both the single position and center-aligned length.
\textbf{Position (unaligned)} condition, whose segments lack a shared start-point reference, showed further degradation.
This progression from single position to unaligned comparison demonstrates how perceptual accuracy deteriorates systematically as reference frames become less stable.

\textbf{Area} exhibits nearly the poorest baseline accuracy among all channels, yet power-law correction with $\alpha = 0.424$ (95\% bootstrap CI $[0.393, 0.490]$; 1{,}000 participant-level resamples, percentile method) yields the largest improvement across all channels (0.531 log-units), substantially narrowing the gap with position and length.

Color channels show negligible improvement under power-law fitting: \textbf{saturation} and \textbf{luminance} improved by only 0.042 and 0.032, respectively, an order of magnitude smaller than area's 0.531-unit improvement, indicating more fundamental limitations rather than correctable nonlinearity.

\textbf{Tilt} fell within a pre-specified $\pm 0.2$ log-error equivalence bound relative to single position (TOST: $p = 0.015$; 90\% CI $[-0.162, 0.071]$; $d_z = -0.062$), while \textbf{curvature} performed comparably to the aligned-position condition.
Both show near-linear perception with minimal benefit from power-law fitting.

\subsection{Discussion}

\noindent\textbf{Reference frame stability shapes spatial accuracy.}
The single-position advantage over center-extended length (the $\alpha = 0.759$ overestimation noted above) reflects the cognitive cost of bidirectional estimation, and explains why unidirectional measurement from a stable reference point, as in bar charts, affords a perceptual advantage.
Even perfectly aligned baselines introduce cognitive load relative to single-anchor judgments, with misalignment compounding the penalty.
Because these effects persist with primitive stimuli stripped of axes and gridlines, the perceptual benefits of spatial alignment reflect intrinsic perceptual constraints rather than chart-decoration artifacts.

\vspace{1mm}
\noindent\textbf{Power-law correction for the area channel.}
Area's substantial improvement under power-law correction aligns with Stevens' square-root hypothesis~\cite{powerlaw}, and with classic cartographic work on graduated symbol scaling~\cite{flannery1971relative}.
This pattern is consistent with the theory that observers estimate area by mentally assessing edge lengths~\cite{spence1991displaying} rather than directly perceiving two-dimensional extent.
Our design does not separate genuine area compression from a side-length heuristic, since both predict the observed compression, and participants received no specific training to discourage judging side length rather than area.
This correction, the largest across all channels, suggests that area's poor reputation largely reflects structured, correctable nonlinearity rather than irreducible noise.
The fitted exponent ($\alpha = 0.424$) is close to, though somewhat below, the square-root value ($0.5$) implied by a side-length heuristic, although reported exponents vary with stimulus geometry (e.g., circles vs. squares) and paradigm.
It thus offers a practical calibration value for area encodings.

\vspace{1mm}
\noindent\textbf{Tilt achieves nearly the highest accuracy in primitive stimuli.}
Tilt's accuracy is statistically equivalent to single position (TOST above), contrary to prior chart-level studies~\cite{GraphicalPerception, channelMturk} that ranked angle-based judgments below position and length.
We set the equivalence margin at $\pm 0.2$ log-error because it is directly interpretable on the task's scale and is practically small relative to the larger cross-channel gaps we emphasize.
This discrepancy suggests that chart-level gaps arise largely from competing cues, such as area and arc length in pie charts, that obscure the orientation signal, not from weak angular perception.
When these confounds are removed, tilt's precision within this equivalence margin rivals that of spatial position (Section~\ref{sec:synthesis}), suggesting that chart-level angle underperformance is largely contextual rather than intrinsic to angular perception.

%% file: sections/05_Discriminability.tex
\section{Discriminability}
\label{sec:discriminability}

\subsection{Task Design}
The discriminability task measures the minimum difference required for reliable visual discrimination, known as the just-noticeable difference (JND) across each channel's range.
This captures how sensitive perception is to small changes, which is critical for distinguishing between similar data values in dense visualizations.

In each trial, two stimuli appeared side by side.
The left stimulus displayed a randomly selected value within the target channel's domain, while the right stimulus varied by $\pm10\%$ of that reference value, selected uniformly in $[-10\%, +10\%]$.
This $\pm 10\%$ range spans up to five times the Weber fraction for the tested channels (2--8\% for spatial and color properties~\cite{Teghtsoonian1971, Harrison2014}), ensuring each trial pair includes stimuli from near-threshold to clearly suprathreshold.

We tested six channels: length, tilt, area, luminance, saturation, and curvature.
Position was excluded because, in this relative same/different task, position and length reduce to the same judgment of spatial extent; the reference-frame advantage of position (Section~\ref{sec:accuracy}) applies to absolute estimation, not relative discrimination (see Limitations).

Participants judged whether the two stimuli were ``Same'' or ``Different''.
A total of 150 participants contributed to the Discriminability experiment.
The 105 core participants each completed 10 trials per channel, and an additional 45 participants each completed 32 trials per channel.
This unequal allocation was part of the experimental design, improving coverage of the two-dimensional reference-value $\times$ offset space for contour extraction and aggregated psychometric fitting to estimate JNDs.
All responses were pooled at the trial level before computing response-probability grids, so each trial contributed equally regardless of which participant group generated it; robustness checks for this pooling are reported in Supplementary~\ref{sup:pooling}.

\subsection{Analysis}

To quantify how discriminability varies across each channel's domain, we computed the proportion of ``Same'' responses for each combination of reference value ($x$) and absolute difference ($|\Delta|$), forming a response probability grid.
Contours at levels 10\%, 20\%, 30\%, 40\%, and 50\% (for the ``Same'' response rate) were extracted and visualized, providing a compact summary of the response surface.
Standard Weber and symmetric boundary models do not capture the asymmetric anchoring patterns evident in these contours, consistent with prior evidence that simple Weber-style models can be insufficient for visualization judgments~\cite{kay2015beyond}.
We therefore developed the \textbf{Anchored Harmonic Weber model}, which extends the classical Weber framework with asymmetric boundary-proximity weighting and fits that shared structure across contour levels.
The minimal detectable difference at a given reference value $x$ is modeled as:
\begin{equation}
    |\Delta|(x) = w_0 \cdot \frac{x}{x_{max}} + 1/{(\frac{1}{w_L x} + \frac{1}{w_R (x_{max} - x)})} + \text{offset}_i
    \label{eq:AHWM}
\end{equation}

The first term, $w_0 \cdot \frac{x}{x_{max}}$, captures the baseline Weber fraction.
The second term is a harmonic-mean combination of two sensitivity terms anchored at the minimum ($w_L$, scaling with $x$) and maximum ($w_R$, scaling with $x_{max} - x$) ends of the domain.
Our model ensures proximity-based dominance consistent with the categorical perception of orientations~\cite{tiltaccuracy}, in which the nearer anchor contributes disproportionately to the threshold.
The boundary term vanishes at each endpoint and peaks in the interior, capturing sharper discrimination near either anchor.

Each contour level is allowed an independent vertical offset, but all contours share the same $w_L$, $w_R$, and $w_0$ parameters, ensuring a common underlying shape across contour levels.
For the tilt channel, response contours revealed a qualitative regime change at $90^\circ$ (vertical), where the contour pattern mirrors around the cardinal axis; we therefore fit separate models for $x < 90^\circ$ and $x \geq 90^\circ$.
No other channel exhibited a comparable mid-range discontinuity, so all remaining channels were fit with a single model spanning the full domain.

The endpoint slopes report the rate at which thresholds decline toward each boundary (left: $w_L + w_0/x_{max}$ and right: $w_R - w_0/x_{max}$), decomposing each boundary's sensitivity into anchor-driven ($w_L, w_R$) and baseline Weber ($w_0$) components.

To confirm that this complexity is warranted, we compared our model against three nested alternatives: constant-JND, standard Weber, and symmetric-anchored models.
Our model achieves the lowest BIC across all seven channel conditions ($\Delta\mathrm{BIC} \geq 52$; Supplementary Table~\ref{tab:nested-full}), confirming that asymmetric boundary anchoring captures structure in the discriminability contours beyond simpler accounts.

\begin{figure}
    \centering
    \includegraphics[width=\linewidth, alt={Discriminability contour maps for length, tilt, area, curvature, luminance, and saturation. Each panel plots the minimal detectable difference against the reference value at five response levels with fitted Anchored Harmonic Weber curves. Saturation grows linearly across its range, luminance stays flat and then falls near white, length and area fall steeply near their maxima, and tilt is most sensitive near 0, 90, and 180 degrees.}]{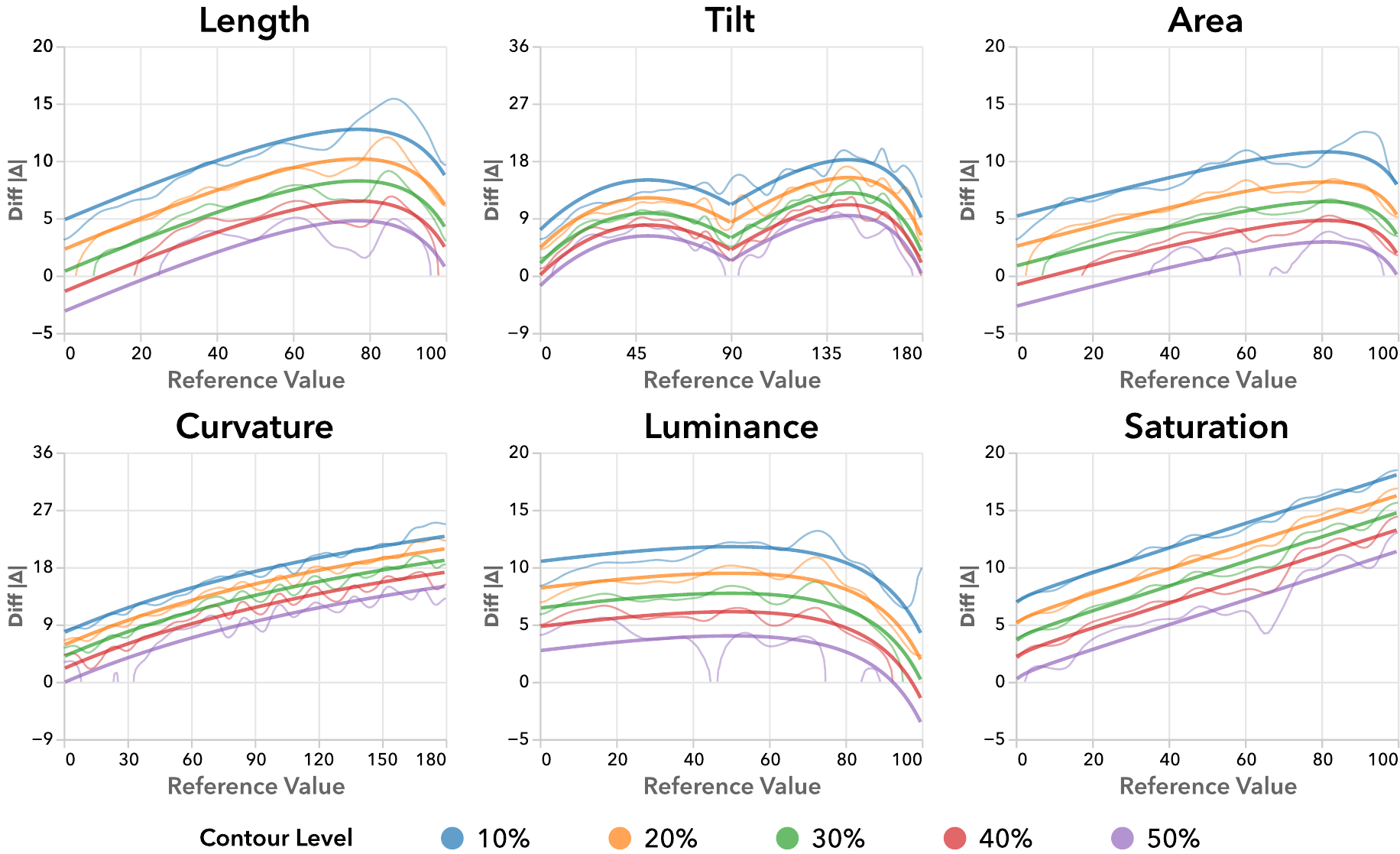}
    \caption{
        Discriminability contour maps for each visual channel.
        The horizontal axis shows the reference value and the vertical axis the minimal detectable difference ($|\Delta|$).
        For each ``Same''-response level (10\%, 20\%, 30\%, 40\%, 50\%, each a distinct color), the thin line is the contour extracted from the response-probability grid and the thick line is the fitted Anchored Harmonic Weber model. A thin line breaks where the grid has no crossing at a given reference value.
    }
    \label{fig:discriminability_contour}
\vspace{-1em}
\end{figure}

\subsection{Results}

Figure~\ref{fig:discriminability_contour} shows the fitted discriminability curves for each channel, estimated from participants' ``Same/Different'' responses across five JND levels.
Channels exhibited diverse discriminability patterns, differing primarily in three respects: (1) how strongly they follow Weber behavior, (2) the number of anchor-like regions shaping the peaks, and (3) boundary effects, where the curve converges to zero toward the endpoints.
Where this convergence is strongest (length and area near their empty end, luminance near its bright end), even the smallest differences we tested were detected by a majority of participants, pushing the fitted 40--50\% contours below the tested range (Supplementary~\ref{sup:nested}).
Fit quality and boundary slopes are summarized in Table~\ref{tab:disc-fit}.

\begin{table}[ht!]
    \caption{Anchored Harmonic Weber model fit ($R^2$) and boundary slopes for each channel.}
    \label{tab:disc-fit}
    \centering
    \begin{tabular}{lccc}
        \toprule
        Channel & $R^2$ & Left slope & Right slope \\
        \midrule
       Area         & 0.93 & 0.0920  & 0.4175   \\
        Curvature    & 0.96 & 0.1898  & $-0.0615$ \\
        Length       & 0.89 & 0.1225  & 2.0544   \\
        Luminance    & 0.93 & 0.0545  & 1.4228   \\
        Saturation   & 0.99 & 0.1279  & $-0.0990$ \\
        Tilt$(0^\circ$--$90^\circ)$     & 0.83 & 0.1204  & 1.0296   \\
        Tilt$(90^\circ$--$180^\circ)$   & 0.87 & 0.1490  & 0.9858   \\
        \bottomrule
    \end{tabular}
\vspace{-1em}
\end{table}

\vspace{2mm}
\textbf{Saturation} best approximates Weber's law pattern in our dataset, with JND increasing linearly across the entire range.
Unlike all other channels, saturation's thresholds grow proportionally through the full range; the negative right slope confirms the absence of boundary anchoring, indicating that maximum saturation does not function as a perceptual reference.
This near-perfect Weber adherence is consistent with ratio-based sensitivity without strong boundary anchors, making saturation the most uniform channel in our discriminability dataset.

\textbf{Luminance} departs substantially from Weber's law.
JND increases only marginally across most of the range, indicating a near-constant noise floor, then falls steeply near maximum luminance.
This pattern is consistent with a fixed threshold dominating the darker range, while boundary-related sensitivity improves discrimination only near white.

\textbf{Tilt} exhibits the most complex pattern, with three distinct high-sensitivity regions creating a segmented discriminability landscape.
We fit separate models for $0^\circ$--$90^\circ$ and $90^\circ$--$180^\circ$.
The high-sensitivity regions at $0^\circ$ (horizontal), $90^\circ$ (vertical), and $180^\circ$ (horizontal again) create sharp convergence points where discriminability improves dramatically, consistent with prior work on cardinal orientations.

\textbf{Area} shows dual anchoring with remarkably asymmetric slopes.
JND decreases toward both empty (0\%) and full (100\%), but the approach to empty is gradual while the approach to full is steep.
This asymmetry indicates that maximum area functions as a stronger boundary reference than empty space, as evidenced by the higher right slope.

\textbf{Length} displays the strongest right-side anchoring in our dataset, with dramatic improvement in discriminability as length approaches maximum.
This extreme right-anchoring suggests that the canvas boundary provides a strong reference for length judgments.

\textbf{Curvature} follows Weber's law through most of its range but departs from it near $180^\circ$.
As the arc approaches a semicircle, the discriminability curve bends sharply upward, then plateaus, suggesting a perceptual boundary near the transition from ``curved line'' to ``half circle.''

\begin{figure}
    \centering
    \includegraphics[width=\linewidth, alt={Bump chart of discriminability rank across the normalized stimulus range. Area and length rank finest at low values, luminance rises from the bottom ranks to rank 1 near the upper boundary, and curvature and saturation decline steadily.}]{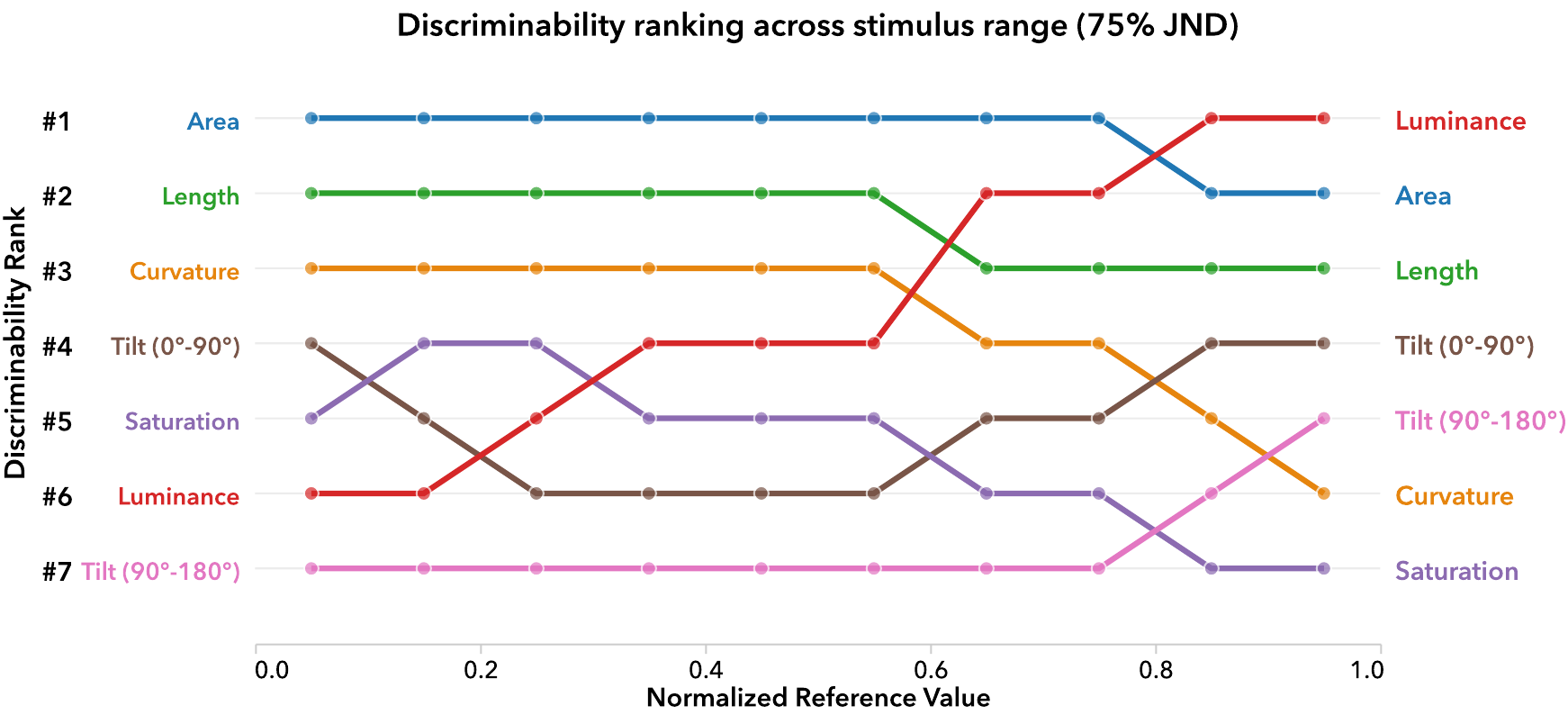}
    \caption{Discriminability ranking across the stimulus range at the 75\% JND threshold.
    Each line tracks a channel's rank (1~=~finest discriminability) as the normalized reference value increases.
    Rankings shift substantially: area and length dominate at low values, but luminance rises from the bottom ranks to rank~1 near the upper boundary, overtaking all other channels.}
    \label{fig:75JND_bump}
\vspace{-1em}
\end{figure}

The 25\% ``Same''-response contour (i.e., 75\% of responses indicated ``Different''), extracted in the same manner, provides a standardized comparison across channels.
Figure~\ref{fig:75JND_bump} tracks each channel's discriminability rank at this threshold across the stimulus range.
At low reference values, area and length achieve the finest discriminability, with luminance and tilt ranking lowest.
This hierarchy then shifts substantially: luminance rises from the bottom ranks to rank~1 near the upper domain boundary, overtaking all other channels, while curvature and saturation degrade steadily.
This dynamic reordering demonstrates that no single discriminability ranking characterizes channel performance across the full stimulus range.

\subsection{Discussion}

\noindent\textbf{Luminance and saturation occupy opposite extremes of discriminability regularity.}
Saturation's near-perfect Weber behavior reflects ratio-based processing without strong perceptual references, whereas luminance's flat profile across the darker range suggests absolute-threshold-like processing, where fixed detection limits override proportional scaling~\cite{Livingstone1988, blackwell1946contrast}.
For design, saturation encodings can span the full range with predictable resolution, whereas luminance encodings should favor the bright end where discriminability is sharpest.

\vspace{1mm}
\noindent\textbf{Tilt's discriminability reflects a segmented perceptual landscape.}
Tilt yields the lowest model fit among all channels ($R^2 = 0.83$/$0.87$), reflecting not measurement noise but a discriminability landscape more complex than a single model can capture.
The cardinal orientations ($0^\circ$, $90^\circ$, $180^\circ$) act as high-sensitivity anchors, necessitating separate half-range models.
Within each half-range, midrange variability remains elevated, consistent with the oblique effect, in which discrimination is poorer at oblique orientations than at cardinal ones~\cite{tiltaccuracy}.
The two half-ranges are not mirror images: discrimination sharpens more steeply toward $90^\circ$ (from below) and $180^\circ$ than toward the opposite ends, a left--right asymmetry we detail in Supplementary~\ref{sup:disc}.
This segmented landscape coexists with tilt's position-level accuracy (Section~\ref{sec:accuracy}), demonstrating that average magnitude estimation and local sensitivity patterns are independent aspects of channel performance.

\noindent\textbf{Maximum extent provides a strong perceptual boundary.}
By \emph{boundary anchoring} we mean that a domain endpoint acts as a perceptual reference, so discrimination sharpens (the contour converges toward zero) as values approach it.
Both length and area exhibit pronounced right-skew in their anchoring profiles.
Length's right slope (2.054) far exceeds its left slope (0.123), and area shows a comparable pattern (right: 0.418 vs.\ left: 0.092).
This pattern suggests that maximum extent or fullness functions as a stronger perceptual boundary than the corresponding near-empty state.
The canvas edge anchors length judgments, and the fully filled region anchors area judgments.
Combined with the accuracy finding that a fixed anchor improves magnitude estimation (Section~\ref{sec:accuracy}), both results indicate that explicit boundary references enhance perception across multiple tasks (Section~\ref{sec:synthesis}).

%% file: sections/06_Separability.tex
\section{Separability}
\label{sec:separability}

\subsection{Task Design}
The separability task assesses whether changes in one channel (the secondary) interfere with judgments about another (the primary).

The same 105 participants who completed the Accuracy experiment also completed the Separability experiment.
For each tested primary--secondary channel pairing (Table~\ref{tab:sep-pairs}), each participant performed a \textit{magnitude-estimation judgment} (the same task as in Section~\ref{sec:accuracy}) on the primary channel under two conditions:
(i)~the secondary fixed at its default (the Section~\ref{sec:accuracy} trials), and
(ii)~the secondary sampled uniformly across its full range (color channels varied in HSL).
This added one response per pairing per participant, enabling within-subjects comparison of fixed versus varying secondary-channel effects.
Rows in Table~\ref{tab:sep-pairs} denote the primary channel; columns denote the secondary channel.
Position was tested only as a secondary channel; as a primary, it would largely replicate the length conditions given their shared spatial extent basis (Section~\ref{sec:accuracy}).      

\begin{table}[t]
    \centering
    \caption{
        Channel pairs included in the separability test.
        A check mark (\cmark) indicates a tested pair, with the row channel as primary and the column channel as secondary.
        Blank cells were excluded either because the combination was geometrically infeasible (e.g., area--curvature, area--length conditions) or because they fell outside the prioritized scope of theoretically motivated pairs (spatial--color, within-color, and geometric--geometric interactions).
    }
    \resizebox{0.48\textwidth}{!}{
        \begin{tabular}{rccccccc}
            \toprule
            \diagbox[width=2cm]{\small Primary}{\small Secondary} & Position & Length & Tilt & Area & Luminance & Saturation & Curvature \\
            \midrule
            Length     & \cmark & & \cmark & & \cmark & \cmark & \cmark \\
            Tilt       & & \cmark & & \cmark & \cmark & \cmark & \cmark \\
            Area       & \cmark & & \cmark & & \cmark & \cmark & \\
            Luminance  & & \cmark & \cmark & \cmark & & \cmark & \cmark \\
            Saturation & & \cmark & \cmark & \cmark & \cmark & & \cmark \\
            Curvature  & & \cmark & \cmark & & \cmark & \cmark & \\
            \bottomrule
        \end{tabular}
    }
    \label{tab:sep-pairs}
\end{table}

\subsection{Analysis}

To quantify separability, we computed per-trial log-error using (Equation~\ref{eq:mean_log_error}) with $\alpha{=}1$, where $P$ and $S$ denote the participant's raw response and the true value, both normalized by the channel maximum $S_{\max}$.

Separability was assessed by the mean log-error difference between the varying and fixed secondary conditions: larger differences indicate stronger interference.
We also tested the statistical significance of the difference relative to the baseline using paired $t$-tests with Benjamini--Hochberg FDR correction applied within each primary channel (FDR level 0.05), with the participant as the unit of analysis.

\subsection{Results}

Table~\ref{tab:separability_results} presents the tested separability matrix.
Each cell reports the mean log-error (Equation~\ref{eq:mean_log_error}, $\alpha{=}1$, normalized to $[0,1]$ and thus negative) for the row (primary) channel judged while the column (secondary) channel varies.
In each row, the gray cell with the underlined value is that row's \emph{baseline}, sitting in its own-channel column (e.g., Length's baseline is in the Length column) and equal to its Section~\ref{sec:accuracy} accuracy; the remaining colored cells report performance while the column (secondary) channel varies.
Dashed cells (\,--\,) are excluded pairs (Table~\ref{tab:sep-pairs}).
Cell color encodes the deviation from that baseline, $\Delta\text{error}=\text{cell}-\text{baseline}$ (blue: improvement, red: degradation, intensity $\propto|\Delta\text{error}|$ within each direction).
Asterisks denote raw paired $t$-test significance, and bold cells survive BH--FDR correction.

The most pronounced effect is the \textbf{tilt}--\textbf{area} interaction.
When area varies as a secondary channel, tilt judgment accuracy degrades from its baseline of $-2.240$ to $-1.342$ ($p < 0.001$, Hedges' $g = 1.285$), the largest effect across all tested pairings, bringing tilt performance near the chance level ($\approx -1.33$, the expected log-error under uniform random responding).
The reversed pairing (area judged as tilt varies) produces a smaller yet significant effect ($-1.685$ to $-1.524$; $p = 0.010$, Hedges' $g = 0.276$).
Tilt also degrades substantially when paired with curvature ($-2.240$ to $-1.843$; $p < 0.001$, $g = 0.579$).

Among color channels, \textbf{luminance} variation disrupts \textbf{saturation} judgments ($-1.843$ to $-1.545$; $p < 0.01$, Hedges' $g = 0.487$), whereas saturation variation has minimal impact on luminance.
Luminance judgments instead improve in several pairings when secondary channels vary (Table~\ref{tab:separability_results}).
Saturation degrades across most tested pairings.

\textbf{Length} judgments remain close to baseline across all tested secondary channels.
Position variation improved length performance ($-2.086$ to $-2.253$; $p < 0.01$, $g = 0.299$).

\newcommand{\sig}[2]{#1\rlap{\textsuperscript{\tiny#2}}}
\newcolumntype{C}{>{\centering\arraybackslash}p{1.9cm}}
\definecolor{celltext}{HTML}{222222}
\definecolor{cellbase}{HTML}{EEEEEE}
\definecolor{celldashcolor}{HTML}{CCCCCC}
\newcommand{\celldash}{\cellcolor{cellbase}\textcolor{celldashcolor}{--}}
\newcommand{\basecell}[1]{\cellcolor{cellbase}\underline{#1}}

\begin{table*}[ht!]
    \centering
    \caption{Channel separability log-error matrix (mean).
    In each row, the gray cell with the \underline{underlined} value marks that row's own-channel baseline (its Section~\ref{sec:accuracy} accuracy); the other cells report performance while the column (secondary) channel varies.
    \textcolor[HTML]{1F77B4}{Blue}/\textcolor[HTML]{C0392B}{red} shading shows improvement/degradation vs.\ baseline (intensity $\propto$ magnitude within each direction).
    Asterisks mark raw paired $t$-test significance (* $p<0.05$, ** $p<0.01$, *** $p<0.001$); \textbf{bold} cells survive Benjamini--Hochberg FDR correction.}
    \setlength{\tabcolsep}{3pt}
    \setlength{\extrarowheight}{2pt}
    {
\color{celltext}
\begin{tabular}{l|CCCCCCC}
  \toprule
  \multirow{2}{*}{Primary Channel}
    & \multicolumn{7}{c}{Secondary Channel} \\[1pt]
  \cline{2-8}
  \rule{0pt}{2.6ex} & Position & Length & Tilt & Area & Luminance & Saturation & Curvature \\
  \midrule

  Length
    & \cellcolor[HTML]{1D5C9A}\textcolor{white}{\textbf{\sig{$-$2.253}{**}}}
    & \basecell{$-$2.086}
    & \cellcolor[HTML]{DAE9F1}$-$2.115
    & \celldash
    & \cellcolor[HTML]{F9E8DF}$-$2.001
    & \cellcolor[HTML]{E7EEF2}$-$2.101
    & \cellcolor[HTML]{EFF0F0}$-$2.090 \\

  Tilt
    & \celldash
    & \cellcolor[HTML]{FAE1D3}$-$2.110
    & \basecell{$-$2.240}
    & \cellcolor[HTML]{68001F}\textcolor{white}{\textbf{\sig{$-$1.342}{***}}}
    & \cellcolor[HTML]{FAE1D3}$-$2.111
    & \cellcolor[HTML]{FAE5D9}$-$2.134
    & \cellcolor[HTML]{EC9678}\textbf{\sig{$-$1.843}{***}} \\

  Area
    & \cellcolor[HTML]{E6EEF2}$-$1.702
    & \celldash
    & \cellcolor[HTML]{FBDBCA}\textbf{\sig{$-$1.524}{*}}
    & \basecell{$-$1.685}
    & \cellcolor[HTML]{F9E8DF}$-$1.600
    & \cellcolor[HTML]{F5EEE9}$-$1.651
    & \celldash \\

  Luminance
    & \celldash
    & \cellcolor[HTML]{17518E}\textcolor{white}{\textbf{\sig{$-$1.765}{*}}}
    & \cellcolor[HTML]{16508C}\textcolor{white}{\textbf{\sig{$-$1.766}{*}}}
    & \cellcolor[HTML]{053061}\textcolor{white}{\textbf{\sig{$-$1.843}{**}}}
    & \basecell{$-$1.590}
    & \cellcolor[HTML]{D6E7F0}$-$1.622
    & \cellcolor[HTML]{D2E5EF}$-$1.626 \\

  Saturation
    & \celldash
    & \cellcolor[HTML]{F8E9E0}$-$1.766
    & \cellcolor[HTML]{F3EFEC}$-$1.827
    & \cellcolor[HTML]{FAE3D5}$-$1.722
    & \cellcolor[HTML]{F6B79A}\textbf{\sig{$-$1.545}{**}}
    & \basecell{$-$1.843}
    & \cellcolor[HTML]{FAE0D2}$-$1.710 \\

  Curvature
    & \celldash
    & \cellcolor[HTML]{F7ECE6}$-$1.970
    & \cellcolor[HTML]{FAE5D9}$-$1.913
    & \celldash
    & \cellcolor[HTML]{E2ECF2}$-$2.040
    & \cellcolor[HTML]{F9E7DD}$-$1.927
    & \basecell{$-$2.019} \\
  \bottomrule
\end{tabular}
}
    \label{tab:separability_results}
\vspace{-1em}
\end{table*}

\subsection{Discussion}

\noindent\textbf{Three-tier interference hierarchy.}
Effect magnitudes reveal three interference tiers across the tested pairings.

\begin{itemize}
    \item \textit{Minimal observed interference.}
    Length with color channels shows the smallest deviations from baseline (within 0.085 log units), reflecting low interference rather than proven independence.
    \item \textit{Moderate observed interference.}
    Curvature occupies an intermediate position, with log-errors ranging from $-1.913$ to $-2.040$ across tested partners (baseline: $-2.019$), indicating curvature is moderately robust to secondary-channel variation.
    \item \textit{Strong interference.}
    Beyond the tilt--area case, the tilt--curvature and saturation--luminance pairings also produce significant performance degradation, with the latter exhibiting pronounced asymmetry.
    Position--length is a notable exception, where the varying position mark improves rather than degrades length judgments, indicating that spatial co-variation adds reference information.
\end{itemize}
\vspace{1mm}
\noindent\textbf{Quantifying channel separability.}
While Garner's framework~\cite{Garner1974} established a clear boundary between separable and integral dimensions, our results reveal a continuum from mild to severe interference.
The tilt--area pair exemplifies one extreme, where area variation fundamentally disrupts tilt processing.
The observed asymmetry (e.g., luminance disrupts saturation but not vice versa) further challenges bidirectional separability assumptions, consistent with hierarchical visual processing in which earlier-processed luminance information interferes with higher-level color judgments~\cite{Livingstone1988}.

We also observed facilitative effects.
Luminance judgments improved with area variation, which likely reflects stimulus-level signal enhancement (larger area increases luminance extent) rather than a cross-channel mechanism; separating these accounts requires independent control of stimulus extent and encoded area value.
The position--length facilitation, by contrast, allows a cleaner interpretation: the varying position mark provides an additional spatial reference, paralleling our accuracy finding that fixed reference frames improve length judgments (Section~\ref{sec:accuracy}).

%% file: sections/07_Popout.tex
\section{Pop-out}
\label{sec:popout}

\subsection{Task Design}
The pop-out task measures preattentive detection under brief exposure.
In each trial, participants viewed an array of 20 elements (e.g., lines) on the screen, with or without a single outlier that differed in exactly one channel.
Beyond the seven core channels, we additionally tested hue and shape (square, triangle, hexagon), which are relevant for pop-out but were excluded from the other three experiments because they lack ordinal magnitude scales.

Before each trial, the target channel was cued: participants were told which channel to monitor during the countdown, so search was guided rather than open-ended. Distractors were circles in the shape and hue conditions and the channel's default mark otherwise; array layout, distractor/outlier construction, and the resulting upper-bound interpretation appear in Supplementary~\ref{sup:popout}.
Following a 3-second countdown, the stimulus array was displayed for 100\,ms, after which participants indicated whether an outlier was present and, if so, confirmed its location.
We measured detection accuracy but not response time.
The 100\,ms exposure is consistent with established preattentive processing paradigms~\cite{Treisman1980, Bridges2020}.
Outlier values were standardized across the magnitude channels: distractors at 25\% and outliers at 75\% of the channel range (or vice versa), yielding a uniform separation of 50\% of the channel range that exceeds the measured JND for every channel we tested in the discriminability task (Section~\ref{sec:discriminability}).
Outlier presence was randomized at 50\%.
We collected data from 105 participants with three trials per channel.
Following singleton-search designs, our paradigm makes two deliberate simplifications: participants knew the target channel in advance, and all distractors were identical~\cite{muller1995visual, chen2015singleton}.

\begin{figure}[t!]
    \centering
    \includegraphics[width=\linewidth, alt={Dot plot of pop-out detection accuracy under 100 ms exposure. Area and hue rank highest, followed by position; luminance and saturation are substantially lower. Among shapes, square and triangle remain salient while hexagon falls to near-chance detection.}]{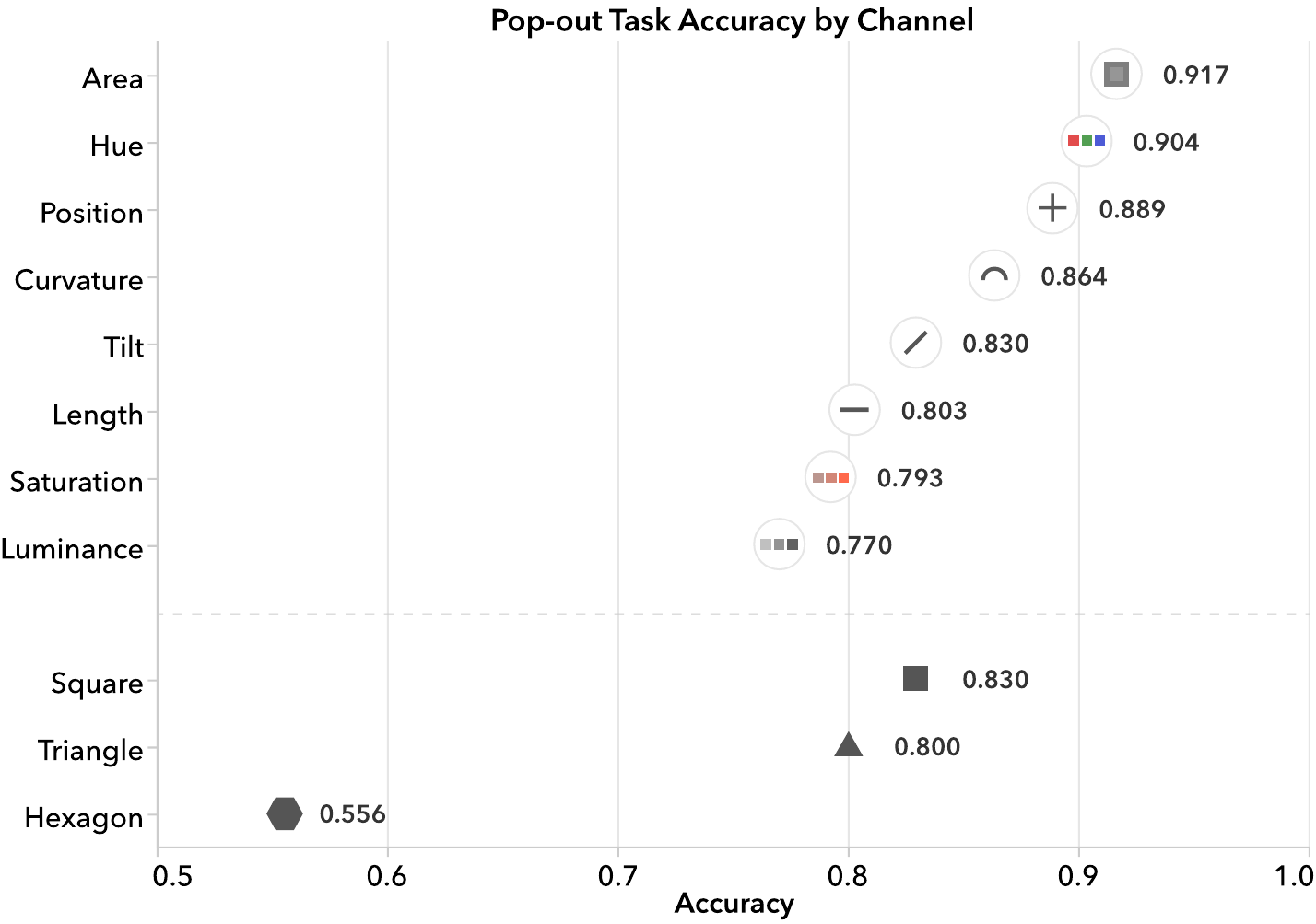}
    \caption{Pop-out detection accuracy across visual channels under 100\,ms exposure.
    Area and hue achieve the highest pop-out rates, while length shows only moderate pop-out effectiveness.
    Among shape channels the hexagon, the form most similar to the circular distractors, falls to near-chance detection, whereas more distinct forms remain salient.
}
    \label{fig:popout_result}
\vspace{-1em}
\end{figure}

\subsection{Results}

Figure~\ref{fig:popout_result} presents pop-out detection accuracy across all tested channels, revealing a clear hierarchy with substantial variation both within and between channel categories.
Among non-shape channels, \textbf{area} achieved the highest detection accuracy.
\textbf{Hue} follows closely, while \textbf{position} also maintains strong performance.
Geometric channels show moderate performance, with \textbf{curvature} and \textbf{tilt} achieving respectable but not exceptional detection rates.

Color channels exhibit striking variability in pop-out effectiveness.
While \textbf{hue} achieves near-optimal performance, \textbf{saturation} and \textbf{luminance} show substantially lower detection rates.
This disparity aligns with color vision research indicating that hue processing engages more fundamental visual pathways than saturation or brightness variations~\cite{Livingstone1988}.
The 0.134 detection gap between hue and luminance represents one of the largest within-category differences observed, suggesting that not all color dimensions are equivalent for pop-out under brief exposure.

\textbf{Length} achieves only moderate pop-out performance despite its exceptional accuracy in quantitative estimation tasks.
Crucially, this mismatch does not depend on area alone: even excluding area, length ranks fifth in detection performance.
Area significantly outperforms length in pop-out detection (paired $t$-test: $p = 0.006$, Hedges' $g = 0.604$); part of this advantage may stem from stimulus extent or luminance cues, which we examine below.

Shape channels revealed a sharp drop from \textbf{square} and \textbf{triangle}, which maintained salience comparable to geometric channels, to \textbf{hexagon}, which fell to near-chance levels (paired $t$-test: $p < 0.001$, Hedges' $g = 1.029$).
This pattern is consistent with the hexagon's close resemblance to the circular distractors; we return to the similarity-versus-complexity question below.
Both contrasts remain significant under Bonferroni correction ($p = 0.012$ and $p < 0.002$) and permutation tests (10{,}000 iterations).

\subsection{Discussion}

\noindent\textbf{Area's detection advantage contradicts accuracy-based expectations.}
Area achieved the highest pop-out detection accuracy despite requiring a power-law correction for quantitative reading.
Two factors may contribute. Area has the finest discriminability (lowest JND) over most of its range (Figure~\ref{fig:75JND_bump}), so the uniform half-range separation is a larger suprathreshold ratio than for other channels. An area outlier also occupies a larger physical footprint, which enhances attentional capture~\cite{proulx2010size} and adds a luminance-contrast cue absent where outlier and distractor extent match.
Our data cannot separate whether area's advantage reflects rapid size processing, low-level luminance contrast, or both.
For visualization design, this finding applies mainly to displays with uniform distractors, where area is a strong channel for anomaly highlighting.
In practice, area often already encodes a data variable (e.g., bubble size), limiting its availability for highlighting.

\vspace{1mm}
\noindent\textbf{Dissociation between accuracy and pop-out.}
Length, by contrast, shows only moderate pop-out despite top-tier accuracy (fifth even when area is excluded), indicating that detection and estimation precision are governed by distinct mechanisms not captured by accuracy-based rankings alone~\cite{GraphicalPerception, channelMturk}.

\vspace{1mm}
\noindent\textbf{Color hierarchy and shape similarity.}
Hue's superior pop-out, plausibly mediated by opponent-color mechanisms~\cite{Livingstone1988}, suggests prioritizing hue over luminance or saturation for highlighting when pop-out is the goal.
The three tested shapes jointly vary geometric complexity and similarity to the circular distractors, so our design cannot separate the two accounts; the hexagon, the polygon most similar to the distractors, was the hardest to detect, suggesting that a more distinctive shape might pop out even if geometrically complex~\cite{gruner2021simple}.
These patterns support Feature Integration Theory while revealing that primitive-stimulus contexts can reduce the effectiveness of features relying on contextual contrast (e.g., tilt) and amplify those that remain effective under brief, guided viewing (e.g., area).

%% file: sections/08_design_implications.tex
\section{General Discussion}
\label{sec:synthesis}

\subsection{A Scenario-driven Perspective on Channel Effectiveness}

Prior studies on visualization effectiveness typically conclude with design implications for specific chart types and tasks, but existing frameworks do not sufficiently account for the \textbf{user's scenario}.
Here, the scenario is distinct from visualization users or tasks such as those taxonomized by Amar et al.~\cite{amar2005low} or Brehmer et al.~\cite{brehmer2013multi}, in that the same abstract task can arise under different circumstances.
Rather than inheriting encoding choices prescribed by visualization type or task category, we encourage designers to break from this convention and proactively identify the channel that best fits their scenario.
Based on our findings, we organize the design implications into four recurring dimensions of scenario (Task Stakes, Data Granularity, Background Interference, and Response Time), each mapped to a corresponding task (accuracy, discriminability, separability, and pop-out).

\textbf{Task Stakes} corresponds to the \textit{accuracy} task, reflecting how precisely a value must be extracted from an encoded channel.
A battery indicator in mobile phones tolerates coarse readout, whereas a scientific figure or finance dashboard may require much more precise extraction.
In low-stakes contexts, a designer may therefore favor lower-accuracy channels for aesthetics or space efficiency, reserving high-accuracy channels, like position along a common scale, for contexts where precise readout is critical.

\textbf{Data Granularity} maps to \textit{discriminability}, capturing how finely the data must be expressed and interpreted by the viewer.
When data is left-skewed with values concentrated at the higher end of the range, saturation may be a poor choice, as it offers lower discriminability in that region.
One mitigation strategy is selective sampling: for the Tilt channel spanning $0^\circ$--$180^\circ$, sampling near the cardinal orientations ($0^\circ$, $90^\circ$, $180^\circ$) exploits the angles where discriminability is sharpest.
Conversely, right-skewed data (such as a Pareto distribution) maps naturally onto the lower range, where saturation is more discriminable.

\textbf{Background Interference} arises when a visual channel must remain perceptible against competing visual elements, which relates to \textit{separability} and becomes especially relevant when charts are embellished to enhance engagement.
For instance, encoding data values through color hue becomes problematic when the background itself is richly colored or textured, a common occurrence in sports infographics or editorial data journalism, whereas position or length channels tend to remain separable under heavier visual embellishment.
Because our separability experiment measures pairwise channel interference, not background decoration directly, we treat this dimension as a design extrapolation from that pairwise evidence rather than a tested claim; the principle is that concurrent visual variation degrades target-channel judgment.

\textbf{Response Time} governs how quickly a value must be retrieved or acknowledged, which relates to \textit{pop-out}.
Consider a digital car dashboard where vehicle speed is rendered as a large numeral at the center.
Though the speed value itself is not encoded through a conventional perceptual channel, its relative importance is communicated through size, causing it to pop out from the surrounding display before the driver consciously searches for it.
A similar principle applies to alert systems and real-time monitoring dashboards, where color, particularly red, can serve as a preattentive signal that draws the viewer's attention to a critical threshold even before any deliberate search begins.

Because rankings shift across tasks and a scenario is shaped by a broad space of design, data, user, and environmental factors, our scenario-driven perspective provides a practical, non-prescriptive guide that enables designers to make principled channel selections grounded in the conditions under which a visualization is actually consumed.

\subsection{Convergence and Divergence with Chart-Level Evidence}

We compare our results with published chart-level studies to distinguish convergences from context-dependent divergences.

\vspace{-0.5em}
\subsubsection{Convergences}

Several key findings align with chart-level evidence.
The tilt--area interference parallels studies showing that area dominates angle in pie charts~\cite{skau2016arcs, Kosara2019}, and the luminance--saturation asymmetry mirrors directional separability in scatterplots~\cite{SmartSzafir2019, Schloss2019}.
The reference-frame advantage also helps explain the accuracy gap between aligned and stacked bar charts~\cite{Talbot2014, channelMturk}, while area's power-law correction replicates value underestimation in treemaps~\cite{kong2010perceptual}.
Finally, the cross-task ranking shifts align with prior evidence of task-dependent channel effectiveness~\cite{McColeman2022, Saket2019}.

\subsubsection{Divergences}

\noindent\textbf{Angle perception in chart contexts.}
Tilt falls within the pre-specified equivalence margin ($\pm 0.2$ log-error) of single position in our primitive stimuli, yet ranks among the worst in pie chart studies~\cite{GraphicalPerception, channelMturk}.
This divergence arises from \emph{channel confusion}: pie charts co-vary angle, area, and arc length, and viewers rely on the more salient area cue~\cite{skau2016arcs, Kosara2019}.
Our result captures isolated orientation judgment, distinct from the angular extent of pie wedges, while chart-level studies capture the \emph{effective} channel actually used in multi-cue displays.
Therefore, angle encodings should not be dismissed as inherently imprecise, but designers must ensure that competing cues do not override the intended channel.

\vspace{1mm}
\noindent\textbf{Pop-out under distractor heterogeneity.}
Area achieves the highest pop-out detection in our paradigm (0.917), where all non-target elements are identical.
However, preattentive detection degrades with heterogeneous distractors~\cite{Haroz2012}.
Our measurements, therefore, represent an upper bound, strongest when non-target marks are uniform (small-multiple designs) and weakest in dense, heterogeneous displays.

\vspace{1mm}
\noindent\textbf{Scaffolding amplifies spatial channels.}
Gridlines and axes disproportionately benefit spatial channels~\cite{GraphicalPerception}, so the accuracy gap between spatial and non-spatial channels is \emph{wider} in scaffolded contexts than in our primitive stimuli.
Conversely, in scaffolding-free contexts (e.g., augmented reality~\cite{Fonnet2019}, embedded visualizations~\cite{willett2017embedded}, data physicalizations~\cite{Dragicevic2021}), our measurements may better predict channel performance than chart-level rankings.

\vspace{1mm}
These comparisons suggest that primitive stimulus and chart-level paradigms are \emph{complementary}: convergences (tilt--area interference, baseline effects, area's power law) point to shared perceptual structure, while divergences occur precisely where theory predicts, namely competing cues (tilt in pie charts), heterogeneous distractors (pop-out), and spatial scaffolding (gridlines for position).
Recognizing these three modulating factors enables designers to predict \emph{when} our findings transfer directly (e.g., scaffolding-free AR or embedded visualizations) and \emph{when} context-specific adjustments are necessary.

\subsection{Limitations}

Our study relied on participants recruited via Mechanical Turk, introducing uncontrolled variation in display hardware, ambient lighting, and screen calibration.
Attention checks mitigated this, though residual noise in color-based channels is possible.

Also, our experiments prioritized breadth over depth, using sparse per-participant trial counts (two responses for accuracy, one per pairing for separability, and three for pop-out).
Evaluating channels across four perceptual tasks and many channel combinations limited how finely we could probe each domain, particularly for separability, display density, and parametric variation in channel magnitudes.
This breadth-first design yields broad multi-task profiles through large group-level aggregation ($N=105$), but lacks individual-level reliability estimation. Higher trial counts per participant would be needed in follow-up studies to characterize individual differences~\cite{Davis2023Risks}.
Sparse per-participant data inflate the standard error of paired differences, so our tests are conservative and the reported effects survive nonetheless.
We randomized task-block order across participants to control for learning and fatigue from the repeated blocks, though residual effects remain possible.

Several findings lack a clear theoretical explanation.
Most notably, luminance showed increased accuracy in all tested separability conditions when paired with an interfering channel.
These effects were small to moderate and their mechanism remains unconfirmed; characterizing them with targeted replication is future work.

The shared canvas frame was a deliberate constant across stimuli, and its influence is most visible in the strong right-boundary anchoring of length and area in discriminability.
These boundary effects may not transfer to frameless displays.

Finally, we note limitations of each perceptual task.
In terms of accuracy, judgment ranges were defined by channel-specific geometry or by the pre-block reference rather than uniformly by the canvas, and these references provide unequal perceptual support across channels.
For length and area, the canvas boundary can serve as a direct comparison reference, while position is read against its own segment and the color channels rely on the pre-block reference.
The measured accuracy advantage of spatial over color channels partly reflects reference-frame availability rather than intrinsic perceptual superiority alone.

For separability, our design was more sensitive to moderate effects than to small ones, with roughly $N \approx 105$ paired observations per comparison.
Accordingly, channel pairs showing no significant difference should not be interpreted as evidence of separability: small effects may remain undetected after FDR correction, and the ``minimal observed interference'' tier reflects the absence of detected interference rather than confirmed independence.

Position was excluded from discriminability and from the separability primary channel under the assumption that side-by-side comparison neutralizes the reference-frame advantage distinguishing position from length in absolute estimation (Section~\ref{sec:accuracy}, $p = 0.009$), consistent with distinctions between absolute identification and relative judgment tasks~\cite{stewart2005absolute}.
This assumption is plausible where both reduce to spatial-extent comparison under simultaneous viewing, but was not tested; future work should verify it by including position in discriminability.

Saturation was tested with a single base hue (red, hue $=0^\circ$); generalization to other hues requires further investigation, as saturation discrimination may vary across the hue circle.
We also tested all channels within normalized ranges (0--100 for metric channels and $0^\circ$--$180^\circ$ for tilt and curvature); perceptual behavior may differ when values are sampled from narrower subranges or different portions.

%% file: sections/09_result.tex
\section{Conclusion}

Our paper presents a systematic evaluation of visual channels across four critical perceptual tasks using primitive visual stimuli that remove chart-specific scaffolding while retaining a shared display frame.

Our findings demonstrate that channel effectiveness is fundamentally multi-dimensional; channels that excel in one task may falter in another, and tested pairwise interactions can override individual channel strengths.
Rather than proposing another unified ranking, our work provides multi-task profiles, separability constraints, and cross-paradigm comparisons, and organizes these findings through a scenario-driven perspective around four dimensions of visualization usage (task stakes, data granularity, background interference, and response time) for context-sensitive channel selection.

Practically, interference-based constraints such as avoiding tilt--area combinations are among our most robust findings and serve as concrete design warnings within the tested setting.
Task-specific channel selection that uses high-salience channels for preattentive detection and high-accuracy channels for precise reading offers a principled alternative to one-size-fits-all guidelines.
These principles can also complement automated recommendation systems~\cite{DBLP:journals/tvcg/WongsuphasawatMS16, automating} through multi-dimensional channel profiles rather than single-metric rankings.

We also suggest several directions for expanding our approach.
Systematic validation with richer stimuli would clarify the boundary conditions of our findings, particularly for the divergences identified in Section~\ref{sec:synthesis}.
Controlled follow-up on facilitative effects (e.g., position--length improvement) may reveal deliberate channel pairings that enhance rather than merely preserve perception.
Finally, extending the separability framework to three-way interactions and individual differences could enable personalized visualization recommendations and reader-adaptive interpretation support~\cite{choe25enhancing}.
As visualization moves beyond traditional charts to more complex forms, such as augmented reality, embedded graphics, and data physicalizations, these fundamental channel-level insights become increasingly critical for effective visualization design.

%% file: sections/10_supplementary.tex
\newpage

\section*{Supplementary Material}
\label{sec:supplementary}

\setcounter{subsection}{0}
\setcounter{subsubsection}{0}
\renewcommand{\thesubsection}{S\arabic{subsection}}
\renewcommand{\thesubsubsection}{\thesubsection.\arabic{subsubsection}}
\setcounter{figure}{0}
\setcounter{table}{0}
\renewcommand{\thefigure}{S\arabic{figure}}
\renewcommand{\thetable}{S\arabic{table}}

\begin{table*}[b!]
    \caption{Nested model comparison for discriminability.
    For each channel, four models of increasing complexity are compared.
    \textbf{Bold} indicates the best (lowest) BIC per channel.
    $k$: total parameters; $n$: data points; RMSE: root mean squared error.}
    \label{tab:nested-full}
    \centering
    \begin{tabular}{ll rr rrrr}
        \toprule
        Channel & Model & $k$ & $n$ & RSS & RMSE & $R^2$ & BIC \\
        \midrule
        \multirow{4}{*}{Area}
            & Null           & 5 & 537 & 1618.81 & 1.736 & 0.517 & 623.98 \\
            & Weber          & 6 & 537 &  579.01 & 1.038 & 0.827 &  78.16 \\
            & Symmetric      & 7 & 537 &  253.51 & 0.687 & 0.924 & $-$359.08 \\
            & \textbf{Full anchored} & \textbf{8} & \textbf{537} & \textbf{227.33} & \textbf{0.651} & \textbf{0.932} & $\mathbf{-411.31}$ \\
        \midrule
        \multirow{4}{*}{Curvature}
            & Null           & 5 & 1082 & 20991.73 & 4.405 & 0.159 & 3243.41 \\
            & Weber          & 6 & 1082 &  3430.45 & 1.781 & 0.863 & 1290.42 \\
            & Symmetric      & 7 & 1082 &  3241.88 & 1.731 & 0.870 & 1236.23 \\
            & \textbf{Full anchored} & \textbf{8} & \textbf{1082} & \textbf{1071.17} & \textbf{0.995} & \textbf{0.956} & $\mathbf{45.01}$ \\
        \midrule
        \multirow{4}{*}{Length}
            & Null           & 5 & 575 & 3781.84 & 2.565 & 0.353 & 1114.84 \\
            & Weber          & 6 & 575 & 1798.18 & 1.768 & 0.692 &  693.72 \\
            & Symmetric      & 7 & 575 &  923.30 & 1.267 & 0.842 &  316.79 \\
            & \textbf{Full anchored} & \textbf{8} & \textbf{575} & \textbf{650.61} & \textbf{1.064} & \textbf{0.888} & $\mathbf{121.87}$ \\
        \midrule
        \multirow{4}{*}{Luminance}
            & Null           & 5 & 571 & 1874.90 & 1.812 & 0.454 &  710.60 \\
            & Weber          & 6 & 571 & 1874.88 & 1.812 & 0.454 &  716.96 \\
            & Symmetric      & 7 & 571 &  930.92 & 1.277 & 0.729 &  323.53 \\
            & \textbf{Full anchored} & \textbf{8} & \textbf{571} & \textbf{229.21} & \textbf{0.634} & \textbf{0.933} & $\mathbf{-470.40}$ \\
        \midrule
        \multirow{4}{*}{Saturation}
            & Null           & 5 & 580 & 6580.23 & 3.368 & 0.181 & 1440.52 \\
            & Weber          & 6 & 580 &  217.34 & 0.612 & 0.973 & $-$531.14 \\
            & Symmetric      & 7 & 580 &  213.05 & 0.606 & 0.974 & $-$536.34 \\
            & \textbf{Full anchored} & \textbf{8} & \textbf{580} & \textbf{104.47} & \textbf{0.424} & \textbf{0.986} & $\mathbf{-943.29}$ \\
        \midrule
        \multirow{4}{*}{Tilt ($0^\circ$--$90^\circ$)}
            & Null           & 5 & 558 & 2731.33 & 2.212 & 0.366 &  917.83 \\
            & Weber          & 6 & 558 & 1560.66 & 1.672 & 0.638 &  611.85 \\
            & Symmetric      & 7 & 558 &  938.50 & 1.297 & 0.782 &  334.39 \\
            & \textbf{Full anchored} & \textbf{8} & \textbf{558} & \textbf{723.10} & \textbf{1.138} & \textbf{0.827} & $\mathbf{195.22}$ \\
        \midrule
        \multirow{4}{*}{Tilt ($90^\circ$--$180^\circ$)}
            & Null           & 5 & 610 & 3565.77 & 2.418 & 0.602 & 1109.13 \\
            & Weber          & 6 & 610 & 3303.59 & 2.327 & 0.632 & 1068.96 \\
            & Symmetric      & 7 & 610 & 1391.62 & 1.510 & 0.845 &  548.00 \\
            & \textbf{Full anchored} & \textbf{8} & \textbf{610} & \textbf{1162.40} & \textbf{1.380} & \textbf{0.867} & $\mathbf{444.62}$ \\
        \bottomrule
    \end{tabular}
\end{table*}

\subsection{Discriminability: Model Comparison and Contour Details}
\label{sup:nested}
\label{sup:disc}

To validate the Anchored Harmonic Weber model (Section~\ref{sec:discriminability}), we compared four nested models of increasing complexity.
Each successive model adds parameters that test a distinct perceptual hypothesis:

\begin{enumerate}
    \item \textbf{Null} (constant JND):
    \[
    |\Delta|(x) = \mathrm{offset}_i
    \]
    Parameters: $\mathrm{offset}_i$ per contour level ($k = 5$, corresponding to the five ``Same''-response thresholds at 10\%--50\%).

    \item \textbf{Weber} (JND proportional to stimulus):
    \[
    |\Delta|(x) = w_0 \cdot \frac{x}{x_{max}} + \mathrm{offset}_i
    \]
    Adds one shared slope $w_0$ ($k = 6$).
    Tests whether discriminability scales with stimulus magnitude.

    \item \textbf{Symmetric anchored} ($w_L = w_R = w$):
    \[
    |\Delta|(x) = w_0 \cdot \frac{x}{x_{max}} + w \cdot \frac{x\,(x_{max} - x)}{x_{max}} + \mathrm{offset}_i
    \]
    Adds one shared boundary weight $w$ ($k = 7$).
    The parabolic term captures improved discriminability near both domain boundaries, but constrains the effect to be symmetric.

    \item \textbf{Full anchored harmonic} (asymmetric $w_L \neq w_R$):
    \[
    |\Delta|(x) = w_0 \cdot \frac{x}{x_{max}} + 1/{\left(\frac{1}{w_L x} + \frac{1}{w_R (x_{max} - x)}\right)} + \mathrm{offset}_i
    \]
    
    Shared parameters $w_0, w_L, w_R$ ($k = 8$).
    Allows each boundary to exert a different anchoring strength.
\end{enumerate}

All models share the same contour-level offsets ($\mathrm{offset}_i$, $i = 1, \ldots, 5$); the models differ only in their shared shape parameters.
For each model, we minimized the sum of squared residuals (RSS) against the extracted discriminability contours and computed $\mathrm{BIC} = n \ln(RSS/n) + k \ln n$, where $n$ is the number of data points and $k$ is the total parameter count.

Table~\ref{tab:nested-full} reports the full comparison.
Across all six channels (with tilt split into two subranges, yielding seven fitted conditions), the full anchored harmonic model achieves the lowest BIC ($\Delta\mathrm{BIC} \geq 52$ relative to the next-best model), confirming that asymmetric dual-boundary anchoring captures perceptual structure beyond what simpler models explain.

\textbf{Contour extraction.}
The fitted contours in Figure~\ref{fig:discriminability_contour} are iso-probability contours of each channel's response-probability grid (the proportion of ``Same'' responses over the reference-value~$\times$~$|\Delta|$ space), to which the Anchored Harmonic Weber model is fit. Where the fitted model dips below zero near a domain boundary, this reflects extrapolation of the parametric fit rather than a negative threshold, as the JND itself is non-negative.

\textbf{Tilt left--right asymmetry.}
We fit the half-ranges $0$--$90^\circ$ and $90$--$180^\circ$ separately because the cardinal orientations act as high-sensitivity anchors.
Within each half the fitted right-end slope exceeds the left-end slope ($0$--$90^\circ$: $1.030$ vs.\ $0.120$; $90$--$180^\circ$: $0.986$ vs.\ $0.149$; Table~\ref{tab:disc-fit}): discrimination sharpens steeply approaching $90^\circ$ from below and $180^\circ$, yet only weakly approaching $90^\circ$ from above, so the $90^\circ$ anchor is asymmetric by approach direction.
We report this as an empirical pattern: orientation was sampled uniformly and the design was not built to dissect the two sides of the $90^\circ$ anchor, so denser sampling near the cardinal axes is needed to confirm it.

\subsection{Stimulus Examples}
\label{sup:stimuli}

\begin{figure*}[ht!]
    \centering
    \includegraphics[width=\textwidth, alt={Twelve representative stimuli: single position (horizontal and vertical), aligned and unaligned position comparison, length, tilt, area, curvature, luminance, and saturation marks, plus shape and hue pop-out arrays in which one outlier appears among red circles.}]{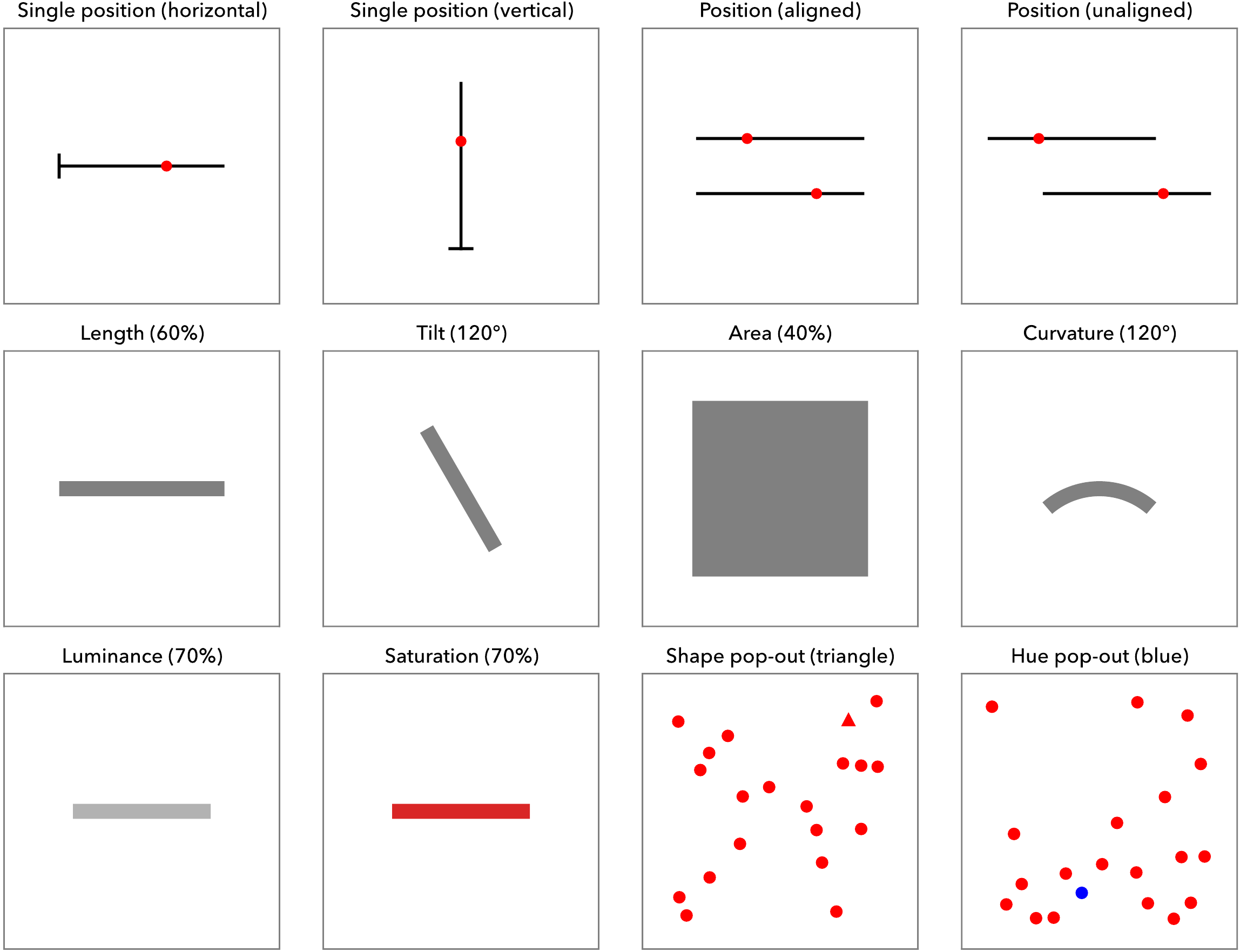}
    \caption{Representative stimuli for each channel and condition.
    Top row: the three position variants (single, aligned, unaligned); single is shown in both horizontal and vertical orientations, and the red dot marks the point to be read.
    Middle row: length, tilt, area, and curvature.
    Bottom row: the achromatic luminance line and the red saturation line (both rendered as colored line marks, not filled squares), and the pop-out arrays for shape (20 elements, an outlier triangle among identical red circles) and hue (an outlier blue circle among red).
    Marks are enlarged slightly for print visibility; per-channel value ranges are in Section~\ref{sec:stimuli_design}.}
    \label{fig:stim}
\end{figure*}

Figure~\ref{fig:stim} shows a representative stimulus for each channel and condition; per-channel value ranges and increments are given in Section~\ref{sec:stimuli_design}, and all marks sit on the 500$\times$500\,px canvas.

\subsection{Analysis Details}
\label{sup:analysis}

\subsubsection{The $1/8$ Offset and Its Sensitivity}
\label{sup:offset}
The additive offset $\tfrac{1}{8}=0.125$ inside the logarithm in Equation~\ref{eq:mean_log_error} is adapted from the log-error measure of Cleveland and McGill and prevents the logarithm from diverging when a response exactly matches the ground truth ($|P-S^\alpha|=0$).
With it, a perfectly correct response maps to $\log_2(0.125)=-3$, the best-case floor on the mean-log-error scale, and chance-level responding under a uniform model corresponds to $\approx-1.33$.
To confirm that our conclusions are not an artifact of this choice, we recomputed the per-channel accuracy baselines and the channel ordering under alternative offsets (Table~\ref{tab:offset-sensitivity}).
The ordering is essentially invariant relative to the reference offset $1/8$ (Spearman $\rho=1.00$ for $1/16$ and $0.97$ for $1/4$ and $1/2$; only the adjacent single-position/tilt and aligned-position/curvature pairs swap at the two largest offsets).
The headline single-position vs.\ length comparison remains significant for offsets $1/16$ ($p=0.002$), $1/8$ ($p=0.009$), and $1/4$ ($p=0.039$), but not for the largest tested offset, $1/2$ ($p=0.12$).

\begin{table}[ht!]
    \caption{Offset sensitivity: per-channel accuracy baseline mean log-error $m$ under alternative offsets, in ranked order. The ordering is essentially unchanged (Spearman $\rho \geq 0.97$ vs.\ the $1/8$ reference).}
    \label{tab:offset-sensitivity}
    \centering
    \small
    \begin{tabular}{lcccc}
        \toprule
        Channel & $1/16$ & $1/8$ & $1/4$ & $1/2$ \\
        \midrule
        Position (single)    & $-2.849$ & $-2.240$ & $-1.524$ & $-0.718$ \\
        Tilt                 & $-2.805$ & $-2.240$ & $-1.545$ & $-0.742$ \\
        Length               & $-2.597$ & $-2.086$ & $-1.442$ & $-0.680$ \\
        Position (aligned)   & $-2.533$ & $-2.029$ & $-1.396$ & $-0.646$ \\
        Curvature            & $-2.499$ & $-2.019$ & $-1.400$ & $-0.656$ \\
        Saturation           & $-2.268$ & $-1.843$ & $-1.278$ & $-0.578$ \\
        Position (unaligned) & $-2.221$ & $-1.791$ & $-1.230$ & $-0.540$ \\
        Area                 & $-2.054$ & $-1.685$ & $-1.171$ & $-0.513$ \\
        Luminance            & $-1.942$ & $-1.590$ & $-1.097$ & $-0.460$ \\
        \bottomrule
    \end{tabular}
\end{table}

\subsubsection{Normality and Nonparametric Robustness}
\label{sup:normality}
Shapiro--Wilk tests did not reject normality for $29$ of the $36$ Accuracy paired-difference distributions, and Wilcoxon signed-rank checks agreed with the headline Accuracy comparisons.
Separately, for the bounded, few-trial Pop-out proportions, where normality is least plausible, the two principal contrasts were verified with sign-flip permutation tests (10{,}000 iterations).

\subsubsection{Pooling of the Two Discriminability Groups}
\label{sup:pooling}
Discriminability used two groups (105 core at 10 trials per channel; 45 additional at 32), merged and pooled at the \emph{trial} level: the response-probability grid averages the ``Same'' indicator over all trials in each (reference value, $|\Delta|$) cell, so every trial contributes equally and participants with 32 trials therefore contribute proportionally more observations to a cell than those with 10. The unequal allocation was an intentional design choice: the additional trials densify coverage of the reference-value~$\times$~offset space for contour extraction. A Mann--Whitney comparison of the two groups' per-participant ``Same'' rates found no detectable difference on five of the six channels; saturation was the exception, so its contours rest more heavily on the pooling assumption. As a stronger check, we refit the Anchored Harmonic Weber model on the core sample alone (10 trials per channel) and compared it with the full-sample fit. The channel discriminability ordering is identical across the two (Spearman $\rho = 1.0$) and the per-channel parameters are consistent, which supports the robustness of the reported ordering, though not full equivalence between the two groups.

\subsection{Pop-out: Array Layout and Cueing Scope}
\label{sup:popout}

Figure~\ref{fig:stim} (bottom row) shows the 20-element pop-out arrays for the shape and hue conditions with the outlier present; distractors are identical within an array (in these two conditions, red circles), so the shape condition pits an outlier shape against circular distractors and the hue condition a single blue circle against red ones.
The positions of the 20 elements were scattered pseudo-randomly within the display.
Before each trial a text prompt named the single channel to monitor, shown during the 3-second countdown, so participants knew which feature dimension to attend before the 100\,ms array appeared.
Combined with the uniform-distractor design, this advance cueing makes the task a guided singleton search: it favors detection and represents an upper bound on preattentive salience rather than pure bottom-up pop-out, where the relevant dimension is unknown and distractors are heterogeneous.
Our detection rates should therefore be read as channel-level salience under guided, uniform-distractor conditions; performance in unguided, heterogeneous displays may be lower (Section~\ref{sec:popout}, Section~\ref{sec:synthesis}).